\documentclass[%
 reprint,
 amsmath,amssymb,
 aps,
 pre,
]{revtex4-1}

\usepackage{graphicx}
\usepackage{dcolumn}
\usepackage{bm}

\newcommand{\Pe}{\mathrm{Pe}}

\begin{document}

\preprint{APS/123-QED}

\title{The entropy production of an active particle in a box}

\author{Nitzan Razin}
 \affiliation{Division of Biology and Bioengineering, California Institute of Technology, Pasadena, CA 91125}

\date{\today}

\begin{abstract}
A run-and-tumble particle in a one dimensional box (infinite potential well) is studied. The steady state is analytically solved and analyzed, 
 revealing the emergent length scale of the boundary layer where particles accumulate near the walls. The mesoscopic steady state entropy production rate of the system is derived from coupled Fokker-Planck equations with a linear reaction term, resulting in an exact analytic expression. The entropy production density is shown to peak at the walls. Additionally, the derivative of the entropy production rate peaks at a system size proportional to the length scale of the accumulation boundary layer, suggesting that the behavior of the entropy production rate and its derivatives as a function of the control parameter may signify a qualitative behavior change in the physics of active systems, such as phase transitions.
\end{abstract}
\maketitle

\section{\label{sec:level1}Introduction}
Active matter is composed of particles that self propel by consuming energy from the environment, producing a persistent random motion. 
Due to the motion persistence, active particles accumulate on surfaces even when the particle-surface interaction is purely repulsive, a behavior which can be described as effective attraction \cite{Wittmann2016,Fily2017}.

A minimal model in which this behavior was studied is that of run-and-tumble particles (RTP), where a free particle moves in a straight line for an exponentially distributed random duration (run), and then changes its direction of motion to a new random direction (tumble). This model was originally suggested to describe the swimming of E. coli bacteria \cite{Schnitzer1993, Tailleur2008, Tailleur2009}.
Similar minimal models, which differ in the statistics of the particle speed and change of direction of motion, are active Brownian particles (ABP) \cite{Fily2012}, and active Ornstein-Uhlenbeck particles (AOUP) \cite{Maggi2015, Farage2015, Fodor2018rev}, which display similar surface accumulation.
In free space, RTPs, ABPs and AOUPs all perform a persistent random walk, which is diffusive at long time and length scales \cite{Fodor2018rev}. However, when confined, they display a nonequilibrium steady state density \cite{Tailleur2009, Elgeti2013,Yang2014,Fily2014,Ezhilan2015,Yan2015,Elgeti2015,Elgeti2016,Bechinger2016,Angelani2017,Wagner2017,Malakar2018,Das2018,Caprini2018,Duzgun2018,Sevilla2019}.
Most theoretical studies of confined active particles used simulations and approximate analytical treatment, including those of 2D and 3D RTPs in a channel \cite{Elgeti2015, Ezhilan2015}. Few models, including the 1D RTP with hard walls, which we study here, have an analytic steady state solution \cite{Paksa2016, Razin2017,Angelani2017,Malakar2018}.

While its steady state particle density can be described by an effective potential \cite{Fily2017,Sevilla2019}, the nonequilibrium nature of a confined active particle system can be quantified by its entropy production.
The entropy production rate (EPR, often called "entropy production") is zero in equilibrium and positive when detailed balance is broken, as signified by the existence of probability currents in the system's state space. It thus provides a quantification of the distance of a system from equilibrium \cite{Schnakenberg1976,Gaspard2004JCP, Gaspard2004JSP, Lan2012, Benjamin2010,Li2019, Busiello2019}.

The entropy production of active matter systems has been a recent topic of interest \cite{Ganguly2013, Chaudhuri2014, Fodor2016, Falasco2016,Nardini2017, Mandal2017, Marconi2017,Pietzonka2017,Shankar2018,Speck2018, Caprini2019,Dabelow2019, Szamel2019,Flenner2020,Chaki2018,Chaki2019,gr2020entropy}.
It has been calculated using different frameworks and at different levels of coarse graining, yielding different results \cite{Shankar2018,Mandal2017,Fodor2016,Pietzonka2017,Speck2018}.
The EPR for a single free active particle has been calculated analytically in \cite{Shankar2018}, and the result was shown to depend on whether the motion dynamics is overdamped and whether the self propulsion force is odd or even under time reversal. Most previous calculations of the EPR of interacting active systems have been approximate (such as field theory approaches \cite{Nardini2017,Caballero2020}) or numerical \cite{Nardini2017,Flenner2020,Speck2018}. In the following, we calculate exactly the EPR of an overdamped dry active particle which interacts with confining hard walls, from Fokker-Planck equations.

In this paper, we study the minimal system of a single RTP in a box in one dimension. The model is defined in Section \ref{sec:model_def}. In Section \ref{sec:stst}, we derive and study the limits of its nonequilibrium steady state. In Section \ref{sec:epr}, we develop an expression for the mesoscopic EPR corresponding to dynamics given by a Fokker-Planck equation with linear reaction terms, and use it to obtain an exact analytic result for the EPR of the system. We show that the EPR has maximal slope at a system size proportional to the persistence length of the particle motion. We define an entropy production density and show that it is maximal near the system edges, where the interaction between the particle and the walls causes breaking of time reversal symmetry.

\section{\label{sec:model_def} Model Definition}
A point-like run-and-tumble particle moves at a constant speed $v$, as a result of a constant magnitude self propulsion force acting against friction in an overdamped regime.
With rate $\alpha$, a tumble event occurs and a new random direction of motion is chosen.
Note that in one dimension, since with probability $1/2$ the new direction of motion is equal to the previous one, the rate of change of direction is $\alpha/2$.
In addition to the active motion, the particle diffuses (due to the temperature or another source of white noise) with diffusion coefficient $D$.

In one dimension, a run-and-tumble particle can move in one of two directions: left or right. This allows writing two coupled Fokker-Planck (FP) equations for the density of right and left moving particles at position $x$ and time $t$ - $R(x,t)$ and $L(x,t)$ \cite{Schnitzer1993,Tailleur2008,Tailleur2009}:
\begin{equation} \label{eq:FP_particle_in_box}
\begin{array}{ll}
\partial_t R = D\partial_x^2R - v\partial_x R + \frac{\alpha}{2}(L-R) \equiv -\partial_x J_R+ \frac{\alpha}{2}(L-R)\\
\\
\partial_t L = D\partial_x^2L + v\partial_x L + \frac{\alpha}{2}(R-L) \equiv -\partial_x J_L+ \frac{\alpha}{2}(R-L)
\end{array}
\end{equation}

The probability currents associated with the right / left moving particles are 
\begin{equation} \label{eq:J_def}
\begin{array}{ll}
J_R(x,t) = v R - D\partial_x R  \\
\\
J_L(x,t) = -v L - D\partial_x L
\end{array}
\end{equation}

\begin{figure}[htb]
\centering
\includegraphics[width=0.5\linewidth]{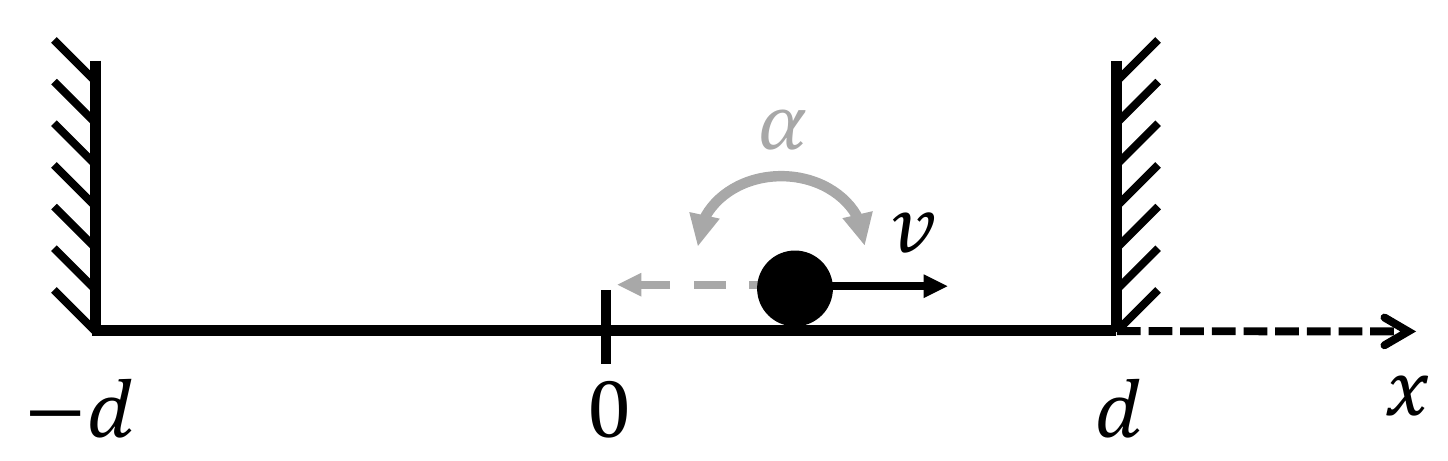}
  \caption[]{System illustration: a run-and-tumble particle (RTP) in a 1D box.}
\label{fig:system_sketch}
\end{figure}

We consider here a RTP in a box - confined by hard walls at $x=\pm d$ (Fig.~\ref{fig:system_sketch}). Thus the particle density is described by Eq.~\ref{eq:FP_particle_in_box}, with reflecting boundary conditions, which mean the currents vanish at the walls: $J_L(\pm d)=J_R(\pm d)=0$.

\section{\label{sec:stst} The Steady State density and current}
We define the total particle density $\rho = R + L$, and the difference $\sigma = R - L$.
The steady state ($\partial_t R = \partial_t L = 0$) solution of Eq.~\ref{eq:FP_particle_in_box}, which satisfies the no flux boundary conditions and the normalization condition of one particle in the box $\int_{-d}^d \! \rho(x) \, \mathrm{d}x=1$ is given by
\begin{equation} \label{eq:rho_sigma_sol}
\begin{array}{ll}
\rho(x) = \rho_0 \cosh(\frac{x}{\xi}) + \rho_1\\
\\
\sigma(x) = \rho_0 \sqrt{1+\Pe^{-1}}\sinh(\frac{x}{\xi})
\end{array}
\end{equation}
where $\xi = \sqrt{\frac{D^2}{v^2+\alpha D}}=\sqrt{\frac{1}{\Pe ( \Pe+1 )}}\ell_p$,  $\Pe=v^2\tau/D$ is the dimensionless Peclet number, $\tau=\alpha^{-1}$ is the persistence time, $\ell_p=v\tau$ is the persistence length, and the constants $\rho_0$ and $\rho_1$ are
\begin{equation} \label{eq:FP_sol_constants}
\begin{array}{ll}
\rho_1 = \frac{1}{2d} - \rho_0 \frac{\xi}{d} \sinh ( \frac{d}{\xi})
\\
\rho_0 = \frac{\Pe}{2 \left( \xi \Pe \sinh ( \frac{d}{\xi}) + d\cosh ( \frac{d}{\xi})\right)}
\end{array}
\end{equation}

Substituting the steady state density Eq.~\ref{eq:rho_sigma_sol} into Eq.~\ref{eq:J_def} gives the currents:
\begin{equation} \label{eq:J_stst_sol}
J_R(x) = \frac{v}{2} \left( -\rho_0 \Pe^{-1} \cosh(\tfrac{x}{\xi}) + \rho_1 \right) = -J_L(x)
\end{equation}

\begin{figure}[!htb]
\centering
\includegraphics[width=1\linewidth]{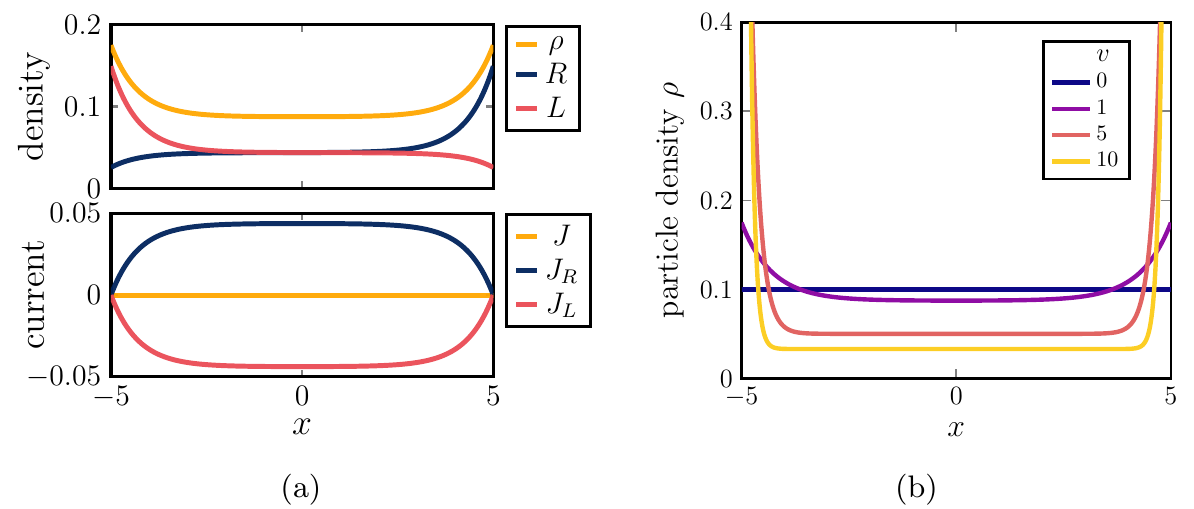}
  \caption[]{The steady state density (Eq.~\ref{eq:rho_sigma_sol}) and current (Eq.~\ref{eq:J_stst_sol}) for $D=1$, $\alpha=1$, $d=5$. (a) Top: The densities of right/left moving particles $R/L$, and their sum - the total density $\rho$. Bottom: The current of right/left moving particles $J_{R/L}$, and their sum - the total current $J=J_R+J_L=0$. ($v=1$) (b) Particle density $\rho(x)$, for varying particle speed $v$. For a Brownian particle ($v=0$), the density profile is uniform. As the active velocity increases, a growing portion of the density is concentrated near the walls.}
\label{fig:system_stst_density_current}
\end{figure}

The particle spends time near the walls due to its persistence: when the particle's active force pushes it against a wall, it takes on average a time duration $2\tau$ to turn around. This results in a boundary region of increased density near each wall, which decays exponentially with length scale $\xi$ (Fig.~\ref{fig:system_stst_density_current}).
The boundary layer width $\xi$ is proportional to the persistence length of the particle motion $\ell_p$, with a dimensionless proportionality constant which is a function of the Peclet number.
The currents of R and L particles vanish at the walls as required by the boundary conditions, and grow in magnitude towards the system center (Fig.~\ref{fig:system_stst_density_current}(a)). Their sum, the total particle current in the system, vanishes ($J=J_R+J_L=0$).

A similar solution appears in \cite{Malakar2018}, and for the $D=0$ case in \cite{Paksa2016,Razin2017,Angelani2017}. The steady state density of the RTP in the box is notably similar to the  approximate solution for the density of ABPs in a 2D channel with hard walls derived in \cite{Yan2015,Duzgun2018}. 

In the limit of vanishing Peclet number $\Pe \to 0$, the thermal equilibrium result of uniform density $\rho(x)=1/2d$, which is the Boltzmann distribution inside the box, is restored.
The length scale of the surface accumulation in the $\Pe \ll 1$ limit is $\xi \simeq \sqrt{D \tau}$, while in the $\Pe \gg 1$ limit, $\xi \simeq D/v$. The accumulation length scale $\xi$ increases with the diffusion coefficient $D$, because diffusion causes wall-facing particles to spread over a nonzero length near the wall, instead of remaining at the wall. 
Without diffusion $D=0$, the length scale of the accumulation boundary region vanishes $\xi = 0$, and the particle density at the walls diverges as a macroscopic number of particles accumulate on each of the walls. This can be shown either by taking the limit of $D \to 0$ in the steady state solution above, or by writing coupled differential equations for the densities of left and right moving particles in the bulk and the numbers of particles accumulated on the walls. Both methods yield the same result (Appendix \ref{appendix:rho_limits}).
Working in the $D=0$ limit is often useful since it allows an analytical solution of some generalizations of the run-and-tumble model, such as ones with position dependent $v$ and $\alpha$ and source and sink terms \cite{Razin2017, Razin2017b}. Moreover, this limit is relevant for colloids and bacteria, where thermal diffusion is typically two orders of magnitude smaller than the effective active diffusion $v^2 \tau$ \cite{Marchetti2016}.

\section{\label{sec:epr} The entropy production}
The steady state density of a particle in a box indicates that the system is out of equilibrium, since it is not the Boltzmann distribution, which is a uniform distribution inside the box.
Nevertheless, one can define an effective energy $E_{\text{eff}}(x)=-\beta^{-1} \log(\rho(x))$, for which the steady state particle distribution is the Boltzmann distribution $\rho(x)\propto \exp(-\beta E_{\text{eff}}(x))$. The effective interaction potential between the particle and the walls is attractive. This effective attraction between the particle and the wall is similar to the effective attraction between self propelled particles with repulsive interactions which causes them to create a dense cluster phase at large enough densities and active speeds in 2 and 3 dimensional systems. This phase transition is known as motility induced phase separation (MIPS) \cite{Cates2015,Farage2015,Marconi2015}.

\begin{figure}[!tb]
\centering
\includegraphics[width=1\linewidth]{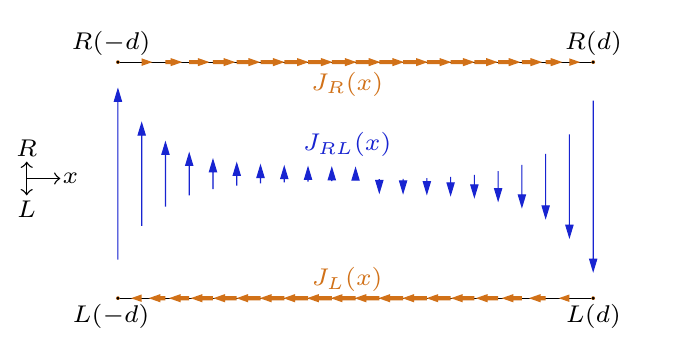}
  \caption[]{Probability flux in state space. Black lines represent the continuum of $R$ and $L$ states for $-d\le x \le d$. There is a flux $J_{R/L}$ between the continuum of $R/L$ states (orange arrows), and a flux density $J_{RL}(x)$ of discrete transitions from the state $L(x)$ to the state $R(x)$ for each $-d\le x \le d$ (blue arrows). The existence of probability flux in state space indicates the breaking of detailed balance and a nonzero production of entropy.}
\label{fig:probability_flux_in_state_space}
\end{figure}

Thus it is not possible to determine if the system is out of equilibrium by observing the steady state particle density. The fact that the total current $J=0$ is also consistent with the equilibrium-like picture.
However, when in addition to the particle positions, the direction of the active force is known, $R(x)$ can be distinguished from $L(x)$. This reveals the the existence of currents (Eq.~\ref{eq:J_stst_sol}, Figs.~\ref{fig:system_stst_density_current}a, \ref{fig:probability_flux_in_state_space}) and breaking of detailed balance.
The deviation of the system's steady state from equilibrium, associated with the severity of the violation of detailed balance, is quantified by the entropy production rate.

The Eqs.~\ref{eq:FP_particle_in_box} describe the diffusion and drift of particles of types R and L, along with a reaction that turns R particles into L and vice versa.
In the following, we derive the entropy production rate for such reaction-drift-diffusion FP equations. Similarly to the derivations in \cite{Schnakenberg1976} for a spatially discrete system defined by master equations, and  in \cite{Tome2006} for the FP equation with drift and diffusion terms, we show that the time derivative of the information/Gibbs entropy $S$ is a sum of two terms: $\dot{S}=\Pi-\Phi$. $\Pi$, which can be interpreted as the entropy production rate, is non-negative and vanishes if the system is in detailed balance. $\Phi$ is the entropy flux from the system to the environment \cite{Schnakenberg1976,Tome2006}.
Moreover, we find that the EPR $\Pi$ for a reaction-drift-diffusion FP equation is composed of separate contributions from the drift-diffusion in space and from the reaction.

The system states are $\mathcal{S}=(i,x)$ for $i=R/L$ and $-d \le x \le d$. The information entropy is given by
\begin{equation} \label{eq:gibbs_S}
\begin{array}{ll}

    S(t) = - \sum\limits_{\mathcal{S}} P(\mathcal{S},t) \log P(\mathcal{S},t) \\
    = -\int\limits_{-d}^d \mathrm{d}x R(x,t) \log R(x,t) - \int\limits_{-d}^d \mathrm{d}x L(x,t) \log L(x,t)
\end{array}
\end{equation}
where in the first line the summation over the continuous degree of freedom $x$ denotes an integral, and $P((R/L,x),t) \equiv R/L(x,t)$ as defined before.
We denote $S_R \equiv -\int_{-d}^d \mathrm{d}x R(x,t) \log R(x,t)$. Its time derivative is
\begin{equation}
\dot{S}_R = -\int_{-d}^{d} \mathrm{d}x \partial_t R (\log R - 1)
\end{equation}

By using Eq.~\ref{eq:FP_particle_in_box}, we get
\begin{equation} 
\dot{S}_R = -\int\limits_{-d}^{d} \mathrm{d}x (\partial_x J_R + J_{LR}) (\log R - 1)
\end{equation}
where $J_{LR}(x)=\frac{\alpha}{2}(R(x)-L(x))$ is the flux density from $R(x)$ to $L(x)$.
Using integration by parts and the boundary condition $J_R(\pm d)=0$, 
\begin{equation}\label{eq:S_R_dot}
\dot{S}_R = -\int\limits_{-d}^{d} \mathrm{d}x J_R \partial_x\log R + \int\limits_{-d}^{d} \mathrm{d}x J_{LR} (\log R - 1)
\end{equation}

From the definition of $J_R$, $D \partial_x \log R = v - J_R/R$, and therefore the first term in the right hand side of Eq.~\ref{eq:S_R_dot} is equal to
\begin{equation}
- \frac{v}{D}\int\limits_{-d}^d \mathrm{d}x J_R + \frac{1}{D}\int\limits_{-d}^d \mathrm{d}x \frac{J_R^2}{R} \equiv -\Phi_R + \Pi_R
\end{equation}

We identify $\Pi_R$ and $\Phi_R$ as the entropy production and entropy flux from the system to the environment due to the drift and diffusion of $R(x)$ \cite{Seifert2005,Tome2006}.
Similarly,
\begin{equation} \label{eq:S_L_dot}
\dot{S}_L = -\Phi_L + \Pi_L - \int\limits_{-d}^{d} \mathrm{d}x J_{LR} (\log L - 1)
\end{equation}
where $\Phi_L\equiv -\frac{v}{D}\int_{-d}^{d}\mathrm{d}x J_L$ and $\Pi_L\equiv \frac{1}{D} \int_{-d}^d \mathrm{d}x \frac{J_L^2}{L}$.

Summing Eq.~\ref{eq:S_R_dot} and Eq.~\ref{eq:S_L_dot}, we find
\begin{equation} \label{eq:S_dot}
\dot{S} = \dot{S}_R + \dot{S}_L = \Pi - \Phi
\end{equation}

where $\Pi$ and $\Phi$ are defined as follows:
\begin{equation} \label{eq:EP_components}
\begin{array}{ll}
\Pi \equiv \Pi_R + \Pi_L + \Pi_{RL} \\
\\
= \frac{1}{D}\int\limits_{-d}^d \mathrm{d}x \left(\frac{J_R^2}{R} + \frac{J_L^2}{L} \right) + \frac{\alpha}{2} \int\limits_{-d}^d \mathrm{d}x (R-L) \log (R/L)
\end{array}
\end{equation}

$\Pi$ is the total entropy production rate. It is non negative and thus satisfies the second law of thermodynamics. $\Pi_{R/L}$ is the entropy production due to drift and diffusion of $R/L$ particles, which generate the flux $J_{R/L}$.  It has the same form as the EPR of a drift-diffusion FP equation \cite{Seifert2005,Tome2006}. $\Pi_{RL}$ is given by

\begin{align}
\Pi_{RL} &\equiv \frac{\alpha}{2} \int\limits_{-d}^{d} \mathrm{d}x (R-L) \log (R/L) \\
&= \frac{\alpha}{2} \left(  D_{KL}(R||L) + D_{KL}(L||R) \right)
\end{align}
where $D_{KL}(f||g)$ is the Kullback-Leibler (KL) divergence of the functions $f$ and $g$.
$\Pi_{RL}$ is the entropy production due to the transitions $R(x) \rightleftharpoons L(x)$, according to the entropy production definition for discrete state systems \cite{Schnakenberg1976,Gaspard2004JSP}, integrated over the continuum of contributions from all $-d \le x \le d$. This symmetrized KL divergence quantifies the difference between the distributions $L(x)$ and $R(x)$, and vanishes when they are equal, as happens in equilibrium ($v=0$).

The second term in Eq.~\ref{eq:S_dot} is
\begin{equation}
\Phi \equiv \Phi_R + \Phi_L =  - \frac{v}{D}\int\limits_{-d}^{d} \mathrm{d}x (J_R - J_L) 
\end{equation}
This is the total entropy flux from the system to the environment, composed of symmetric contributions from $R$ and $L$.

In steady state, $\dot{S}=0$ and thus $\Pi=\Phi$. We can therefore calculate the EPR $\Pi$ by evaluating the simpler expression for $\Phi$. We obtain:
\begin{equation} \label{eq:EPR}
\Pi =  \alpha \frac{\frac{d}{\xi} \cosh \left( \frac{d}{\xi} \right) - \sinh \left( \frac{d}{\xi} \right)}{\Pe^{-1} \frac{d}{\xi} \cosh \left( \frac{d}{\xi} \right) + \sinh \left( \frac{d}{\xi} \right)}
\end{equation}

In the equilibrium limit $\Pe=0$, the entropy production vanishes $\Pi=0$. As $\Pe$ grows, the entropy production grows. For $\Pe \gg 1$, $\Pi \simeq \alpha \left(1-d/\xi \coth(d/\xi)\right)$, which diverges for vanishing $\xi$ - as $D\to0$, $v \to \infty$ or $\alpha \to \infty$.
The divergence occurs since in this limit, there are wall-facing particles accumulated on the walls. These particles have a nonzero rate of transition to the state with opposite propulsion force direction, while the reverse probability flux vanishes since the number of particles at the wall with propulsion force away from it is zero (Appendix \ref{appendix:rho_limits}).

The EPR depends on the system half length $d$ and the accumulation length scale $\xi$ only through their dimensionless ratio $d/\xi$. The EPR vanishes in the small system limit $d/\xi \to 0$, and monotonically increases as $d$ is increased.
In the infinitely large system $d \to \infty$ limit, $\Pi \to v^2/D$  which is the entropy production of a free particle (as calculated in \cite{Shankar2018} for overdamped, time reversal symmetry even propulsion).

The rate of change of the EPR as the system size is varied, $\partial_d\Pi$, is maximal for $d_{\text{max}}=\xi f(\Pe)$, where $f(\Pe)$ is a constant close to one for $\Pe \le 1$, and grows logarithmically with $\Pe$ for $\Pe \gg1$ (Appendix \ref{appendix:EP_derivative}). The fact that the EPR slope with respect to $d$ is maximal for a system size proportional to $\xi$ is interesting in light of previous findings in nonequilibrium systems were the EPR or its slope were maximal near (or divergent at) a critical point \cite{Benjamin2010, Gaspard2004JCP,Tome2012}. While this 1D system does not undergo a phase transition, $d \approx \xi$ is a transition point between two regimes in which the system is qualitatively different: at $d\gg \xi$ the system is much larger than the particle persistence length $\ell_p$ and thus in the bulk the particle motion is diffusive at long time and length scales, with diffusion constant $D+v^2\tau$. Once the system size - a length scale introduced by the particle interaction with the walls - becomes of the order of $\xi$, the system is dominated by the boundary physics.

Since the boundary accumulation of the particle on the walls resembles that of particles on each other in MIPS, it would be interesting to see if a similar maximum in the EPR or its slope occurs near the critical volume in 2 and 3 dimensional systems of interacting active particles that undergo MIPS.

\begin{figure}[!tb]
\centering
\includegraphics[width=1\linewidth]{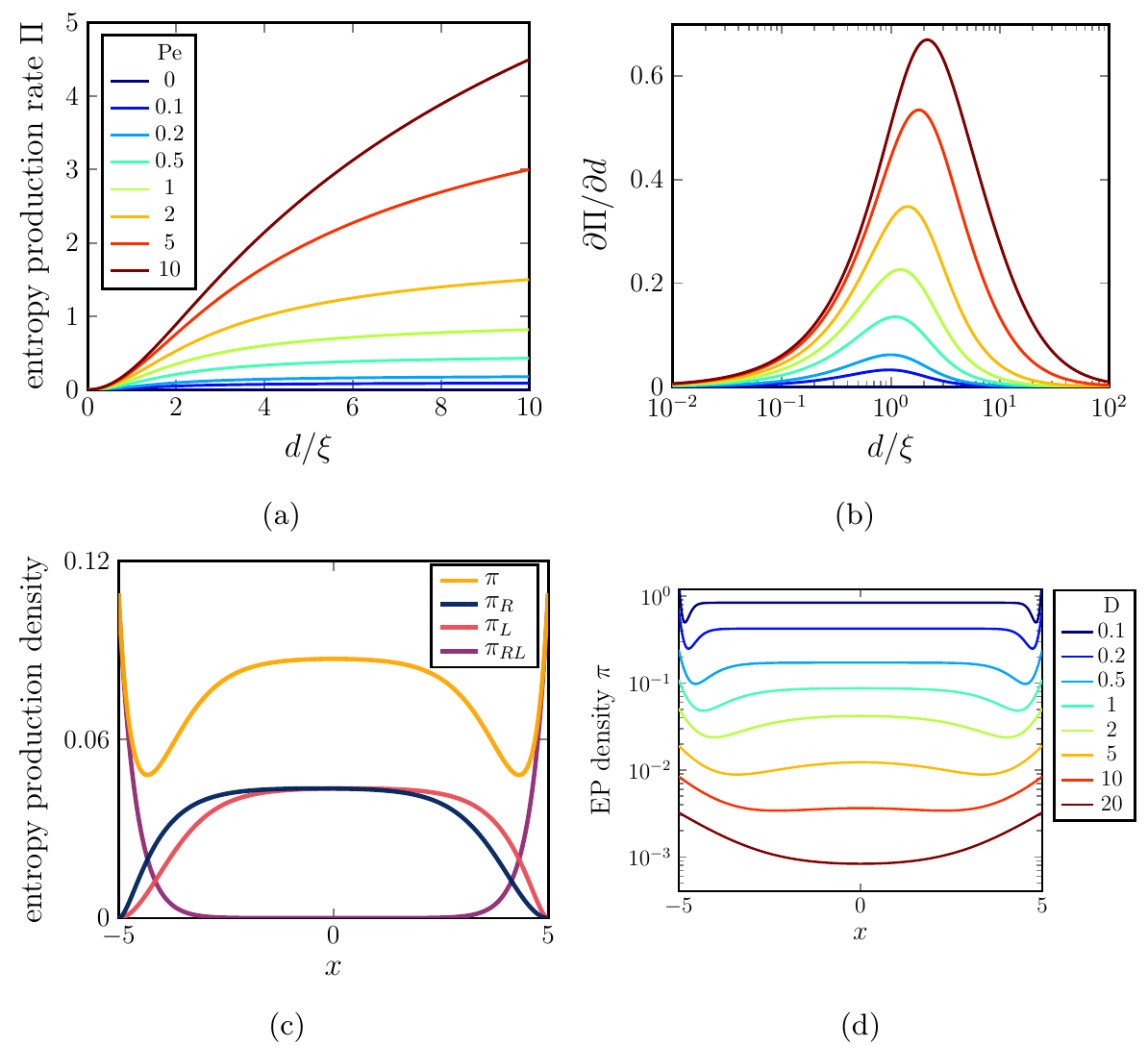}
  \caption[]{Entropy production of a RTP in a 1D box. (a) The entropy production rate $\Pi$ (Eq.~\ref{eq:EPR}) and (b) $\partial_{d} \Pi$ as a function of $d/\xi$, for varying $\Pe$ values ($\alpha=1$). The derivative $\partial_{d} \Pi$ is maximal at $d_{\text{max}}/\xi=f(\Pe)$ (independent of $\alpha$ and $\xi$), and decays to zero away from $d_{\text{max}}$.  (c) The entropy production density and its three components (Eq.~\ref{eq:EP_density_components}) for $v=1$, $\alpha=1$, $D=1$, $d=5$. It is maximal at the walls. (d) The EPR density $\pi(x)$ with $v,\alpha,d$ as in c and varying $D$. There is a qualitative change in the shape of $\pi(x)$ as the diffusion coefficient $D$ varies.}
\label{fig:entropy_production}
\end{figure}

From Eq.~\ref{eq:EP_components}, the entropy production is given by an integral over the system length of a quantity that we can identify as an entropy production density and denote it by $\pi(x)$. As with the total EPR $\Pi$, we can consider the three separate contributions to the EPR density of the current of the right moving current $J_R$, the current of left moving particle states $J_L$, and the probability current between the states $R(x)$ and $L(x)$ for each $x$, denoting them $\pi_R(x)$, $\pi_L(x)$ and $\pi_{RL}(x)$ respectively:
\begin{equation} \label{eq:EP_density_components}
\begin{array}{ll}
\pi_{R}(x) \equiv \frac{J_R^2}{DR} \vspace{2mm}\\
\pi_{L}(x) \equiv \frac{J_L^2}{DL} \vspace{2mm}\\
\pi_{RL}(x) \equiv \frac{\alpha}{2} (R-L)\log(R/L)
\end{array}
\end{equation}

$\pi_{RL}(x)$ is symmetric, and peaks near the walls, where the difference $R-L$ is the largest. For $\xi \ll d$, it vanishes in the bulk, where $R(x)\approx L(x)$. $\pi_R(x)$ and $\pi_L(x)$ are asymmetric mirror images of each other (as are R and L) which vanish at the walls and peak in the bulk where the current magnitude per particle density is largest. The total entropy production density is concentrated at the walls (Fig.~\ref{fig:entropy_production}c,d), consistent with the fact that time reversal symmetry is violated by the particle interaction with the walls. This resembles the findings of \cite{Nardini2017}, where the entropy production density for a cluster of active particles in the motility induced phase separation regime was found to be maximal at the interface, which could function similarly to a wall for the fluid phase on one of its sides.

\section{Discussion}
A RTP in a 1D box is an analytically solvable minimal system in which the nonequilibrium physics of active particles can be studied at the mesoscopic level described by Fokker-Planck equations. While a single active particle in free space undergoes effective diffusion at large time and length scales, in the case of the infinite potential well the interaction of the particle with the hard walls breaks detailed balance also at the coarse grained level \cite{Tailleur2009}. We showed here that the violation of detailed balance in this system can be quantified by analytically calculating the entropy production rate from a Fokker-Planck description.

We obtained an expression for the entropy production rate in the system which is an integral over local quantities resulting from local detailed balance breaking, and can be interpreted as an entropy production density. This density is maximal near the walls, similarly to the results of \cite{Nardini2017} for interacting particles in 2D, in which the entropy production density was shown to be maximal at the boundary of a particle cluster in system at MIPS.

We found that the EPR derivative as a function of the box size is maximal at a size proportional to the length scale of the boundary region of increased density near the walls.
This result, and similar findings in other nonequilibrium systems \cite{Benjamin2010, Gaspard2004JCP,Tome2012}, suggest that the behavior of the entropy production and its derivatives may provide an interesting characterization of active systems at regimes of qualitative change of behavior, and specifically at criticality. In particular, it would be interesting to measure these quantities as a control parameter, such as the system volume, is varied across the motility induced phase separation transition, since that phenomenon resembles the surface accumulation displayed by a confined active particle: In MIPS, below the critical volume, the system separates into dense clusters and a dilute fluid phase. The  mechanisms for clustering, in which the active particles appear to effectively attract each other, is similar to the effective attraction of the RTP in a 1D box to the walls and results from the persistence of their motion.

Indeed, it was recently found that within a field theory of active phase transitions, the EPR diverges at criticality \cite{Caballero2020}. In addition, \cite{Flenner2020} recently showed that the entropy production rate as defined by \cite{Fodor2016} has a maximum as a function of the persistence time in a system of AOUPs in 3D. In light of our results for the entropy production of a confined RTP, we suggest that the maximum may be a signature of a qualitative transition in the system behavior.

\begin{acknowledgments}
I thank Raphael Voituriez, David Van Valen and Rob Phillips for useful discussions, and Yuval Baum for a critical reading of the manuscript.
\end{acknowledgments}

\appendix

\section{limits of the steady state distribution}\label{appendix:rho_limits}
\subsection{The no diffusion ($D \to 0$) limit}

\subsubsection{Taking the $D \to 0$ limit of the steady state density Eq.~\ref{eq:rho_sigma_sol}}
As $D \to 0$, the length scale of wall accumulation $\xi \to 0$, yet the number of particles accumulated near the wall remains finite - resulting in a diverging particle density.
We first derive an expression for the asymptotic behavior of $\rho(x)$ for small $D$. Since $\Pe \gg 1$, $\xi = \frac{\ell_p}{\Pe} + O(\Pe^{-2})$. Using this and $\xi \ll d$, we get:
\begin{equation}
\rho(x) \simeq \frac{1}{2(d+\ell_p)}\left(1 + \Pe \exp \left(\frac{|x|-d}{\xi} \right) \right)
\end{equation}

Taking the limit $\xi \to 0$, and using the identity $\lim_{\epsilon \to 0} \frac{1}{2 \epsilon} e^{|x|/\epsilon}=\delta(x)$, we obtain:
\begin{equation} \label{eq:rho_D_0}
\lim_{D \to 0} \rho(x) = \frac{1}{2(d+\ell_p)} \left( 1 + 2\ell_p \delta(|x|-d) \right)
\end{equation}
where in order to maintain the particle number conservation normalization condition $\int_{-d}^d\rho(x) \mathrm{d}x = 1$, we define an integral over a delta function with the integration boundary at the point of the delta divergence to be equal to $1/2$: $\int_0^a\delta(x)=\frac{1}{2}$ for $a\neq 0$ (consistent with the contribution of the asymptotic expression for small $\xi$ to the integral).

Thus Eq.~\ref{eq:rho_D_0} describes a constant bulk density $\rho(x) = \frac{1}{2(d+\ell_p)}$ for $-d<x<d$, and a macroscopic particle number accumulated on each of the walls:
\begin{equation}
N^{-d}= \frac{\ell_p}{d+\ell_p} \lim_{\epsilon \to 0} \int\limits_{-d}^{-d+\epsilon} \mathrm{d}x \delta(|x|-d) =  \frac{\ell_p}{2(d+\ell_p)} 
\end{equation}
and similarly the number of particles at $x=d$ is $N^d=N^{-d}$, where we denote the number of particles at $x=\pm d$ by $N^x$.

\subsubsection{Solving coupled equations directly for $D=0$}
The coupled rate equations for the bulk density of left and right moving particles, and for the numbers of particles accumulated on the walls:
\begin{equation} \label{eq:FP_D_0}
\begin{array}{ll}
\vspace{1mm}
\partial_t R = -\partial_x J_R + \frac{\alpha}{2}(L-R) \\
\vspace{1mm}
\partial_t L = -\partial_x J_L + \frac{\alpha}{2}(R-L) \\
\vspace{1mm}
\partial_t N_L^{-d} = -J_L(-d)-\frac{\alpha}{2} N_L^{-d} \\
\vspace{1mm}
\partial_t N_R^{-d} = -J_R(-d)+\frac{\alpha}{2} N_L^{-d} \\
\vspace{1mm}
\partial_t N_L^{d} = J_L(d)+\frac{\alpha}{2} N_R^{d} \\
\partial_t N_R^{d} = J_R(d)-\frac{\alpha}{2} N_R^{d} ,\\
\end{array}
\end{equation}
where $J_R(x,t) = vR(x,t)$ and $J_L = -vL(x,t)$ are the currents of right and left moving particles, and $N_{L/R}^x$ is the number of left/right-moving particles at the boundary position $x=\pm d$. The number of particles accumulated on a wall that are moving away from it is zero (i.e. $N_R^{-d}=N_L^d=0$).
We set the total particle number to $1$: $\int_{-d}^d\rho(x) \mathrm{d}x + N_L^{-d}+N_R^d = 1$, where $\rho(x)=R(x)+L(x)$ is the total particle density. The steady state solution of Eq.~\ref{eq:FP_D_0} is then
\begin{equation} \label{eq:FP_D_0_sol}
\begin{array}{ll}
\vspace{1mm}
\rho(x)\equiv \rho_0 = \frac{1}{2(d+\ell_p)}\\
\vspace{1mm}
R(x)=L(x)=\frac{1}{2}\rho_0 \\
N_L^{-d}=N_R^d=\frac{\ell_p}{2(d+\ell_p)} , \; N_R^{-d}=N_L^d=0
\end{array}
\end{equation}

This is the same as the solution derived in the previous subsection by taking the $D\to 0 $ limit of the steady state density.
More general versions of these equations allowing position dependent $v$ and $\alpha$, and including sink and source terms, have analytic steady state solutions that were studied in \cite{Razin2017, Razin2017b}.

\section{Dependence of the entropy production derivative $\partial_d \Pi$ on $\Pe$} \label{appendix:EP_derivative}

We show below that the entropy production derivative as a function of the system size, $\partial_d\Pi(d; \alpha,\xi,\Pe)$, is maximal at $d_{\text{max}}=\xi f(\Pe)$ where
\begin{equation} \label{eq:f_Pe_asymptotes}
f(\Pe) \simeq 
\begin{cases}
      \mbox{constant} \approx 0.91, & \text{if}\ \Pe\ll1 \\
      -\frac{1}{2}W_{-1}(\frac{1}{2\Pe}), & \text{if}\ \Pe \gg 1
\end{cases}
\end{equation}
where $W_{-1}(x)$ is the $-1$ branch of the Lambert W function.

We defined the model by the FP equations Eq.~\ref{eq:FP_particle_in_box} in terms of 4 independent parameters: the speed $v$, the tumble rate $\alpha$, the diffusion coefficient $D$ and the system half size $d$.
We now study the change in entropy production $\Pi$ as $d$ is varied, as a function of an alternative set of 4 independent parameters, which instead of $D$ and $v$ includes the dimensionless Peclet number $\Pe$, and the accumulation length scale  $\xi$ arising from the steady state density (Eq.~\ref{eq:rho_sigma_sol}).

The derivative of the EPR $\Pi$ with respect to the system half length $d$:

\begin{align} \label{eq:EPR_dd}
\partial_d\Pi & =   \frac{\alpha\Pe(1+\Pe)\left(\sinh(\frac{2d}{\xi})-2\frac{d}{\xi} \right)}{2\xi\left( \frac{d}{\xi}\cosh(\frac{d}{\xi})+\Pe \sinh(\frac{d}{\xi})\right)^2}\nonumber\\ 
& = \frac{\alpha\Pe(1+\Pe)}{2\xi} g(\tilde{d},\Pe)
\end{align}
where we denote $\tilde{d}\equiv \frac{d}{\xi}$.
$\partial_d\Pi$ has a single maximum as a function of d (Fig.~\ref{fig:entropy_production}b). From the form of Eq.~\ref{eq:EPR_dd}, the maximum point $d_{\text{max}}=\underset{d}{\mathrm{argmax}}\partial_d\Pi=\xi \tilde{d}_{\text{max}}$ for $\tilde{d}_{\text{max}}=\underset{\tilde{d}}{\mathrm{argmax}}g(\tilde{d},\Pe)\equiv f(\Pe)$ independent of $\alpha$ and $\xi$.

\begin{figure}[tb]
\centering
\includegraphics[width=1\linewidth]{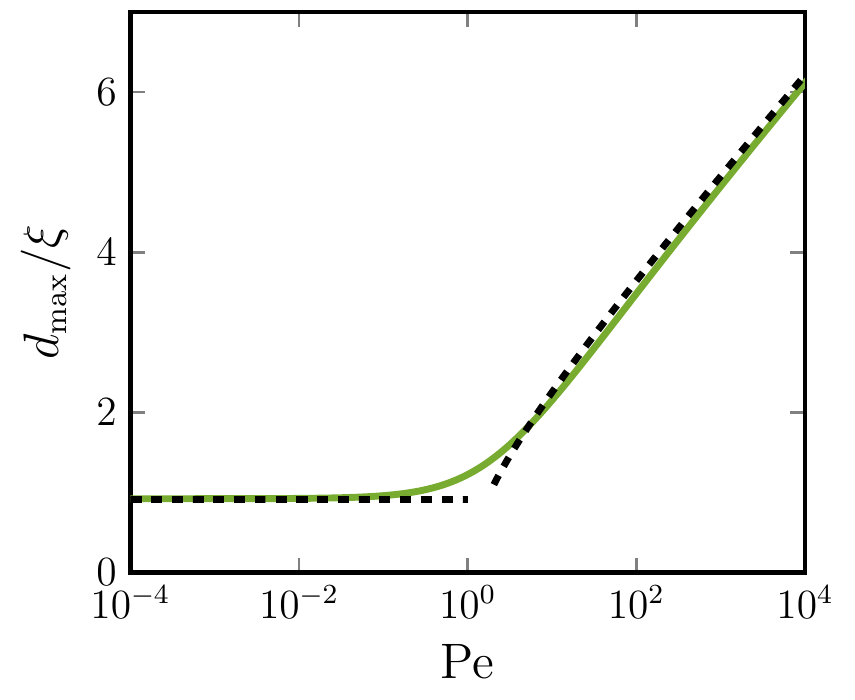}
  \caption[]{The maximum point $d_{\text{max}}$ of the entropy production derivative $\partial_d\Pi(d;\alpha,\xi,\Pe)$ (Eq.~\ref{eq:EPR_dd}). The $\xi$ normalized maximum point $d_{\text{max}}/\xi=f(\Pe)$ is independent of $\alpha$ and $\xi$, and its dependence on $\Pe$ is plotted from numerical calculation (green line). The dashed black lines are analytic results for the asymptotic behavior of $f(\Pe)$ in the $\Pe \ll 1$ and $\Pe \gg 1$ limits (Eq.~\ref{eq:f_Pe_asymptotes}).}
\label{fig:dmax_vs_Pe}
\end{figure}

A numerical evaluation of $\tilde{d}_{\text{max}}$ is plotted in Fig.~\ref{fig:dmax_vs_Pe}. It appears that $f(\Pe)$ has distinct behaviours in the two regimes $\Pe \ll 1$ and $\Pe \gg 1$. We shall perform a self-consistent analytic approximation of $\tilde{d}_{\text{max}}=f(\Pe)$ in each of these regimes.

First, in the $\Pe \ll 1$ limit, it seems that $\tilde{d}_{\text{max}}$ converges to a constant which is slightly smaller than 1. In this limit,
\begin{equation}
g(\tilde{d},\Pe) \simeq	\frac{\left(\sinh(2\tilde{d})-2\tilde{d} \right)}{ \tilde{d}^2\cosh^2(\tilde{d})}
\end{equation}

Taking a 2nd order Taylor expansion around $\tilde{d}=1$ and calculating the maximum of the resulting parabola gives $\tilde{d}_{\text{max}} \approx 0.91$, consistent with the Taylor approximation assumption.

Next, consider the $\Pe \gg 1$ limit. From Fig.~\ref{fig:dmax_vs_Pe}, in this limit $\tilde{d}_{\text{max}}$ grows logarithmically with $\Pe$. Thus we will look for an approximate solution for $\tilde{d}_{\text{max}}$ when $\Pe \gg \tilde{d} \gg 1$. Using the approximation $\sinh(x)\approx \cosh(x) \approx \frac{1}{2}e^x$ for $x\gg1$, we obtain:
\begin{equation}
g(\tilde{d},\Pe) \simeq	\frac{2\left(1-4\tilde{d}e^{-2\tilde{d}} \right)}{\left( \tilde{d}+ \Pe \right)^2}
\end{equation}

Looking for a maximum by demanding that $\partial_{\tilde{d}}g(\tilde{d})=0$, and keeping only leading order terms for $\Pe \gg x \gg 1$, gives the demand $e^{2\tilde{d}}=4\tilde{d}\Pe$. The solution of this equation is
\begin{equation}
\tilde{d}_{\text{max}}=-\frac{1}{2}W_{-1}\left(\frac{1}{2\Pe}\right)
\end{equation}
where $W_{-1}(x)$ is the $-1$ branch of the Lambert W function.
The results of the two approximations of the asymptotic behavior of $\tilde{d}_{\text{max}}$ are plotted in Fig.~\ref{fig:dmax_vs_Pe}, showing that the numerical evaluation of the maximum converges to the approximate results in the respective limits.


\begin{thebibliography}{58}%
\makeatletter
\providecommand \@ifxundefined [1]{%
 \@ifx{#1\undefined}
}%
\providecommand \@ifnum [1]{%
 \ifnum #1\expandafter \@firstoftwo
 \else \expandafter \@secondoftwo
 \fi
}%
\providecommand \@ifx [1]{%
 \ifx #1\expandafter \@firstoftwo
 \else \expandafter \@secondoftwo
 \fi
}%
\providecommand \natexlab [1]{#1}%
\providecommand \enquote  [1]{``#1''}%
\providecommand \bibnamefont  [1]{#1}%
\providecommand \bibfnamefont [1]{#1}%
\providecommand \citenamefont [1]{#1}%
\providecommand \href@noop [0]{\@secondoftwo}%
\providecommand \href [0]{\begingroup \@sanitize@url \@href}%
\providecommand \@href[1]{\@@startlink{#1}\@@href}%
\providecommand \@@href[1]{\endgroup#1\@@endlink}%
\providecommand \@sanitize@url [0]{\catcode `\\12\catcode `\$12\catcode
  `\&12\catcode `\#12\catcode `\^12\catcode `\_12\catcode `\%12\relax}%
\providecommand \@@startlink[1]{}%
\providecommand \@@endlink[0]{}%
\providecommand \url  [0]{\begingroup\@sanitize@url \@url }%
\providecommand \@url [1]{\endgroup\@href {#1}{\urlprefix }}%
\providecommand \urlprefix  [0]{URL }%
\providecommand \Eprint [0]{\href }%
\providecommand \doibase [0]{http://dx.doi.org/}%
\providecommand \selectlanguage [0]{\@gobble}%
\providecommand \bibinfo  [0]{\@secondoftwo}%
\providecommand \bibfield  [0]{\@secondoftwo}%
\providecommand \translation [1]{[#1]}%
\providecommand \BibitemOpen [0]{}%
\providecommand \bibitemStop [0]{}%
\providecommand \bibitemNoStop [0]{.\EOS\space}%
\providecommand \EOS [0]{\spacefactor3000\relax}%
\providecommand \BibitemShut  [1]{\csname bibitem#1\endcsname}%
\let\auto@bib@innerbib\@empty
\bibitem [{\citenamefont {Wittmann}\ and\ \citenamefont
  {Brader}(2016)}]{Wittmann2016}%
  \BibitemOpen
  \bibfield  {author} {\bibinfo {author} {\bibfnamefont {R.}~\bibnamefont
  {Wittmann}}\ and\ \bibinfo {author} {\bibfnamefont {J.~M.}\ \bibnamefont
  {Brader}},\ }\href {\doibase 10.1209/0295-5075/114/68004} {\bibfield
  {journal} {\bibinfo  {journal} {{EPL} (Europhysics Letters)}\ }\textbf
  {\bibinfo {volume} {114}},\ \bibinfo {pages} {68004} (\bibinfo {year}
  {2016})}\BibitemShut {NoStop}%
\bibitem [{\citenamefont {Fily}\ \emph {et~al.}(2017)\citenamefont {Fily},
  \citenamefont {Baskaran},\ and\ \citenamefont {Hagan}}]{Fily2017}%
  \BibitemOpen
  \bibfield  {author} {\bibinfo {author} {\bibfnamefont {Y.}~\bibnamefont
  {Fily}}, \bibinfo {author} {\bibfnamefont {A.}~\bibnamefont {Baskaran}}, \
  and\ \bibinfo {author} {\bibfnamefont {M.~F.}\ \bibnamefont {Hagan}},\ }\href
  {\doibase 10.1140/epje/i2017-11551-3} {\bibfield  {journal} {\bibinfo
  {journal} {The European Physical Journal E}\ }\textbf {\bibinfo {volume}
  {40}},\ \bibinfo {pages} {61} (\bibinfo {year} {2017})}\BibitemShut {NoStop}%
\bibitem [{\citenamefont {Schnitzer}(1993)}]{Schnitzer1993}%
  \BibitemOpen
  \bibfield  {author} {\bibinfo {author} {\bibfnamefont {M.~J.}\ \bibnamefont
  {Schnitzer}},\ }\href {\doibase 10.1103/PhysRevE.48.2553} {\bibfield
  {journal} {\bibinfo  {journal} {Phys. Rev. E}\ }\textbf {\bibinfo {volume}
  {48}},\ \bibinfo {pages} {2553} (\bibinfo {year} {1993})}\BibitemShut
  {NoStop}%
\bibitem [{\citenamefont {Tailleur}\ and\ \citenamefont
  {Cates}(2008)}]{Tailleur2008}%
  \BibitemOpen
  \bibfield  {author} {\bibinfo {author} {\bibfnamefont {J.}~\bibnamefont
  {Tailleur}}\ and\ \bibinfo {author} {\bibfnamefont {M.~E.}\ \bibnamefont
  {Cates}},\ }\href {\doibase 10.1103/PhysRevLett.100.218103} {\bibfield
  {journal} {\bibinfo  {journal} {Phys. Rev. Lett.}\ }\textbf {\bibinfo
  {volume} {100}},\ \bibinfo {pages} {218103} (\bibinfo {year}
  {2008})}\BibitemShut {NoStop}%
\bibitem [{\citenamefont {Tailleur}\ and\ \citenamefont
  {Cates}(2009)}]{Tailleur2009}%
  \BibitemOpen
  \bibfield  {author} {\bibinfo {author} {\bibfnamefont {J.}~\bibnamefont
  {Tailleur}}\ and\ \bibinfo {author} {\bibfnamefont {M.~E.}\ \bibnamefont
  {Cates}},\ }\href {\doibase 10.1209/0295-5075/86/60002} {\bibfield  {journal}
  {\bibinfo  {journal} {EPL (Europhysics Letters)}\ }\textbf {\bibinfo {volume}
  {86}},\ \bibinfo {pages} {60002} (\bibinfo {year} {2009})}\BibitemShut
  {NoStop}%
\bibitem [{\citenamefont {Fily}\ and\ \citenamefont
  {Marchetti}(2012)}]{Fily2012}%
  \BibitemOpen
  \bibfield  {author} {\bibinfo {author} {\bibfnamefont {Y.}~\bibnamefont
  {Fily}}\ and\ \bibinfo {author} {\bibfnamefont {M.~C.}\ \bibnamefont
  {Marchetti}},\ }\href {\doibase 10.1103/PhysRevLett.108.235702} {\bibfield
  {journal} {\bibinfo  {journal} {Phys. Rev. Lett.}\ }\textbf {\bibinfo
  {volume} {108}},\ \bibinfo {pages} {235702} (\bibinfo {year}
  {2012})}\BibitemShut {NoStop}%
\bibitem [{\citenamefont {Maggi}\ \emph {et~al.}(2015)\citenamefont {Maggi},
  \citenamefont {Marconi}, \citenamefont {Gnan},\ and\ \citenamefont
  {Di~Leonardo}}]{Maggi2015}%
  \BibitemOpen
  \bibfield  {author} {\bibinfo {author} {\bibfnamefont {C.}~\bibnamefont
  {Maggi}}, \bibinfo {author} {\bibfnamefont {U.~M.~B.}\ \bibnamefont
  {Marconi}}, \bibinfo {author} {\bibfnamefont {N.}~\bibnamefont {Gnan}}, \
  and\ \bibinfo {author} {\bibfnamefont {R.}~\bibnamefont {Di~Leonardo}},\
  }\href {http://dx.doi.org/10.1038/srep10742} {\bibfield  {journal} {\bibinfo
  {journal} {Scientific Reports}\ }\textbf {\bibinfo {volume} {5}},\ \bibinfo
  {pages} {10742} (\bibinfo {year} {2015})}\BibitemShut {NoStop}%
\bibitem [{\citenamefont {Farage}\ \emph {et~al.}(2015)\citenamefont {Farage},
  \citenamefont {Krinninger},\ and\ \citenamefont {Brader}}]{Farage2015}%
  \BibitemOpen
  \bibfield  {author} {\bibinfo {author} {\bibfnamefont {T.~F.~F.}\
  \bibnamefont {Farage}}, \bibinfo {author} {\bibfnamefont {P.}~\bibnamefont
  {Krinninger}}, \ and\ \bibinfo {author} {\bibfnamefont {J.~M.}\ \bibnamefont
  {Brader}},\ }\href {\doibase 10.1103/PhysRevE.91.042310} {\bibfield
  {journal} {\bibinfo  {journal} {Phys. Rev. E}\ }\textbf {\bibinfo {volume}
  {91}},\ \bibinfo {pages} {042310} (\bibinfo {year} {2015})}\BibitemShut
  {NoStop}%
\bibitem [{\citenamefont {Fodor}\ and\ \citenamefont
  {Marchetti}(2018)}]{Fodor2018rev}%
  \BibitemOpen
  \bibfield  {author} {\bibinfo {author} {\bibfnamefont {{\'E}.}~\bibnamefont
  {Fodor}}\ and\ \bibinfo {author} {\bibfnamefont {M.~C.}\ \bibnamefont
  {Marchetti}},\ }\href {\doibase https://doi.org/10.1016/j.physa.2017.12.137}
  {\bibfield  {journal} {\bibinfo  {journal} {Physica A: Statistical Mechanics
  and its Applications}\ }\textbf {\bibinfo {volume} {504}},\ \bibinfo {pages}
  {106 } (\bibinfo {year} {2018})},\ \bibinfo {note} {lecture Notes of the 14th
  International Summer School on Fundamental Problems in Statistical
  Physics}\BibitemShut {NoStop}%
\bibitem [{\citenamefont {Elgeti}\ and\ \citenamefont
  {Gompper}(2013)}]{Elgeti2013}%
  \BibitemOpen
  \bibfield  {author} {\bibinfo {author} {\bibfnamefont {J.}~\bibnamefont
  {Elgeti}}\ and\ \bibinfo {author} {\bibfnamefont {G.}~\bibnamefont
  {Gompper}},\ }\href {http://stacks.iop.org/0295-5075/101/i=4/a=48003}
  {\bibfield  {journal} {\bibinfo  {journal} {EPL (Europhysics Letters)}\
  }\textbf {\bibinfo {volume} {101}},\ \bibinfo {pages} {48003} (\bibinfo
  {year} {2013})}\BibitemShut {NoStop}%
\bibitem [{\citenamefont {Yang}\ \emph {et~al.}(2014)\citenamefont {Yang},
  \citenamefont {Manning},\ and\ \citenamefont {Marchetti}}]{Yang2014}%
  \BibitemOpen
  \bibfield  {author} {\bibinfo {author} {\bibfnamefont {X.}~\bibnamefont
  {Yang}}, \bibinfo {author} {\bibfnamefont {M.~L.}\ \bibnamefont {Manning}}, \
  and\ \bibinfo {author} {\bibfnamefont {M.~C.}\ \bibnamefont {Marchetti}},\
  }\href {\doibase 10.1039/C4SM00927D} {\bibfield  {journal} {\bibinfo
  {journal} {Soft Matter}\ }\textbf {\bibinfo {volume} {10}},\ \bibinfo {pages}
  {6477} (\bibinfo {year} {2014})}\BibitemShut {NoStop}%
\bibitem [{\citenamefont {Fily}\ \emph {et~al.}(2014)\citenamefont {Fily},
  \citenamefont {Baskaran},\ and\ \citenamefont {Hagan}}]{Fily2014}%
  \BibitemOpen
  \bibfield  {author} {\bibinfo {author} {\bibfnamefont {Y.}~\bibnamefont
  {Fily}}, \bibinfo {author} {\bibfnamefont {A.}~\bibnamefont {Baskaran}}, \
  and\ \bibinfo {author} {\bibfnamefont {M.~F.}\ \bibnamefont {Hagan}},\ }\href
  {\doibase 10.1039/C4SM00975D} {\bibfield  {journal} {\bibinfo  {journal}
  {Soft Matter}\ }\textbf {\bibinfo {volume} {10}},\ \bibinfo {pages} {5609}
  (\bibinfo {year} {2014})}\BibitemShut {NoStop}%
\bibitem [{\citenamefont {Ezhilan}\ \emph {et~al.}(2015)\citenamefont
  {Ezhilan}, \citenamefont {Alonso-Matilla},\ and\ \citenamefont
  {Saintillan}}]{Ezhilan2015}%
  \BibitemOpen
  \bibfield  {author} {\bibinfo {author} {\bibfnamefont {B.}~\bibnamefont
  {Ezhilan}}, \bibinfo {author} {\bibfnamefont {R.}~\bibnamefont
  {Alonso-Matilla}}, \ and\ \bibinfo {author} {\bibfnamefont {D.}~\bibnamefont
  {Saintillan}},\ }\href {\doibase 10.1017/jfm.2015.520} {\bibfield  {journal}
  {\bibinfo  {journal} {Journal of Fluid Mechanics}\ }\textbf {\bibinfo
  {volume} {781}} (\bibinfo {year} {2015}),\ 10.1017/jfm.2015.520}\BibitemShut
  {NoStop}%
\bibitem [{\citenamefont {Yan}\ and\ \citenamefont {Brady}(2015)}]{Yan2015}%
  \BibitemOpen
  \bibfield  {author} {\bibinfo {author} {\bibfnamefont {W.}~\bibnamefont
  {Yan}}\ and\ \bibinfo {author} {\bibfnamefont {J.~F.}\ \bibnamefont
  {Brady}},\ }\href {\doibase 10.1017/jfm.2015.621} {\bibfield  {journal}
  {\bibinfo  {journal} {Journal of Fluid Mechanics}\ }\textbf {\bibinfo
  {volume} {785}} (\bibinfo {year} {2015}),\ 10.1017/jfm.2015.621}\BibitemShut
  {NoStop}%
\bibitem [{\citenamefont {Elgeti}\ and\ \citenamefont
  {Gompper}(2015)}]{Elgeti2015}%
  \BibitemOpen
  \bibfield  {author} {\bibinfo {author} {\bibfnamefont {J.}~\bibnamefont
  {Elgeti}}\ and\ \bibinfo {author} {\bibfnamefont {G.}~\bibnamefont
  {Gompper}},\ }\href {http://stacks.iop.org/0295-5075/109/i=5/a=58003}
  {\bibfield  {journal} {\bibinfo  {journal} {EPL (Europhysics Letters)}\
  }\textbf {\bibinfo {volume} {109}},\ \bibinfo {pages} {58003} (\bibinfo
  {year} {2015})}\BibitemShut {NoStop}%
\bibitem [{\citenamefont {Elgeti}\ and\ \citenamefont
  {Gompper}(2016)}]{Elgeti2016}%
  \BibitemOpen
  \bibfield  {author} {\bibinfo {author} {\bibfnamefont {J.}~\bibnamefont
  {Elgeti}}\ and\ \bibinfo {author} {\bibfnamefont {G.}~\bibnamefont
  {Gompper}},\ }\href {\doibase 10.1140/epjst/e2016-60070-6} {\bibfield
  {journal} {\bibinfo  {journal} {The European Physical Journal Special
  Topics}\ }\textbf {\bibinfo {volume} {225}},\ \bibinfo {pages} {2333}
  (\bibinfo {year} {2016})}\BibitemShut {NoStop}%
\bibitem [{\citenamefont {Bechinger}\ \emph {et~al.}(2016)\citenamefont
  {Bechinger}, \citenamefont {Di~Leonardo}, \citenamefont {L\"owen},
  \citenamefont {Reichhardt}, \citenamefont {Volpe},\ and\ \citenamefont
  {Volpe}}]{Bechinger2016}%
  \BibitemOpen
  \bibfield  {author} {\bibinfo {author} {\bibfnamefont {C.}~\bibnamefont
  {Bechinger}}, \bibinfo {author} {\bibfnamefont {R.}~\bibnamefont
  {Di~Leonardo}}, \bibinfo {author} {\bibfnamefont {H.}~\bibnamefont
  {L\"owen}}, \bibinfo {author} {\bibfnamefont {C.}~\bibnamefont {Reichhardt}},
  \bibinfo {author} {\bibfnamefont {G.}~\bibnamefont {Volpe}}, \ and\ \bibinfo
  {author} {\bibfnamefont {G.}~\bibnamefont {Volpe}},\ }\href {\doibase
  10.1103/RevModPhys.88.045006} {\bibfield  {journal} {\bibinfo  {journal}
  {Rev. Mod. Phys.}\ }\textbf {\bibinfo {volume} {88}},\ \bibinfo {pages}
  {045006} (\bibinfo {year} {2016})}\BibitemShut {NoStop}%
\bibitem [{\citenamefont {Angelani}(2017)}]{Angelani2017}%
  \BibitemOpen
  \bibfield  {author} {\bibinfo {author} {\bibfnamefont {L.}~\bibnamefont
  {Angelani}},\ }\href {\doibase 10.1088/1751-8121/aa734c} {\bibfield
  {journal} {\bibinfo  {journal} {Journal of Physics A: Mathematical and
  Theoretical}\ }\textbf {\bibinfo {volume} {50}},\ \bibinfo {pages} {325601}
  (\bibinfo {year} {2017})}\BibitemShut {NoStop}%
\bibitem [{\citenamefont {Wagner}\ \emph {et~al.}(2017)\citenamefont {Wagner},
  \citenamefont {Hagan},\ and\ \citenamefont {Baskaran}}]{Wagner2017}%
  \BibitemOpen
  \bibfield  {author} {\bibinfo {author} {\bibfnamefont {C.~G.}\ \bibnamefont
  {Wagner}}, \bibinfo {author} {\bibfnamefont {M.~F.}\ \bibnamefont {Hagan}}, \
  and\ \bibinfo {author} {\bibfnamefont {A.}~\bibnamefont {Baskaran}},\ }\href
  {\doibase 10.1088/1742-5468/aa60a8} {\bibfield  {journal} {\bibinfo
  {journal} {Journal of Statistical Mechanics: Theory and Experiment}\ }\textbf
  {\bibinfo {volume} {2017}},\ \bibinfo {pages} {043203} (\bibinfo {year}
  {2017})}\BibitemShut {NoStop}%
\bibitem [{\citenamefont {Malakar}\ \emph {et~al.}(2018)\citenamefont
  {Malakar}, \citenamefont {Jemseena}, \citenamefont {Kundu}, \citenamefont
  {Kumar}, \citenamefont {Sabhapandit}, \citenamefont {Majumdar}, \citenamefont
  {Redner},\ and\ \citenamefont {Dhar}}]{Malakar2018}%
  \BibitemOpen
  \bibfield  {author} {\bibinfo {author} {\bibfnamefont {K.}~\bibnamefont
  {Malakar}}, \bibinfo {author} {\bibfnamefont {V.}~\bibnamefont {Jemseena}},
  \bibinfo {author} {\bibfnamefont {A.}~\bibnamefont {Kundu}}, \bibinfo
  {author} {\bibfnamefont {K.~V.}\ \bibnamefont {Kumar}}, \bibinfo {author}
  {\bibfnamefont {S.}~\bibnamefont {Sabhapandit}}, \bibinfo {author}
  {\bibfnamefont {S.~N.}\ \bibnamefont {Majumdar}}, \bibinfo {author}
  {\bibfnamefont {S.}~\bibnamefont {Redner}}, \ and\ \bibinfo {author}
  {\bibfnamefont {A.}~\bibnamefont {Dhar}},\ }\href {\doibase
  10.1088/1742-5468/aab84f} {\bibfield  {journal} {\bibinfo  {journal} {Journal
  of Statistical Mechanics: Theory and Experiment}\ }\textbf {\bibinfo {volume}
  {2018}},\ \bibinfo {pages} {043215} (\bibinfo {year} {2018})}\BibitemShut
  {NoStop}%
\bibitem [{\citenamefont {Das}\ \emph {et~al.}(2018)\citenamefont {Das},
  \citenamefont {Gompper},\ and\ \citenamefont {Winkler}}]{Das2018}%
  \BibitemOpen
  \bibfield  {author} {\bibinfo {author} {\bibfnamefont {S.}~\bibnamefont
  {Das}}, \bibinfo {author} {\bibfnamefont {G.}~\bibnamefont {Gompper}}, \ and\
  \bibinfo {author} {\bibfnamefont {R.~G.}\ \bibnamefont {Winkler}},\ }\href
  {\doibase 10.1088/1367-2630/aa9d4b} {\bibfield  {journal} {\bibinfo
  {journal} {New Journal of Physics}\ }\textbf {\bibinfo {volume} {20}},\
  \bibinfo {pages} {015001} (\bibinfo {year} {2018})}\BibitemShut {NoStop}%
\bibitem [{\citenamefont {Caprini}\ and\ \citenamefont {Marini
  Bettolo~Marconi}(2018)}]{Caprini2018}%
  \BibitemOpen
  \bibfield  {author} {\bibinfo {author} {\bibfnamefont {L.}~\bibnamefont
  {Caprini}}\ and\ \bibinfo {author} {\bibfnamefont {U.}~\bibnamefont {Marini
  Bettolo~Marconi}},\ }\href {\doibase 10.1039/C8SM01840E} {\bibfield
  {journal} {\bibinfo  {journal} {Soft Matter}\ }\textbf {\bibinfo {volume}
  {14}},\ \bibinfo {pages} {9044} (\bibinfo {year} {2018})}\BibitemShut
  {NoStop}%
\bibitem [{\citenamefont {Duzgun}\ and\ \citenamefont
  {Selinger}(2018)}]{Duzgun2018}%
  \BibitemOpen
  \bibfield  {author} {\bibinfo {author} {\bibfnamefont {A.}~\bibnamefont
  {Duzgun}}\ and\ \bibinfo {author} {\bibfnamefont {J.~V.}\ \bibnamefont
  {Selinger}},\ }\href {\doibase 10.1103/PhysRevE.97.032606} {\bibfield
  {journal} {\bibinfo  {journal} {Phys. Rev. E}\ }\textbf {\bibinfo {volume}
  {97}},\ \bibinfo {pages} {032606} (\bibinfo {year} {2018})}\BibitemShut
  {NoStop}%
\bibitem [{\citenamefont {Sevilla}\ \emph {et~al.}(2019)\citenamefont
  {Sevilla}, \citenamefont {Arzola},\ and\ \citenamefont
  {Cital}}]{Sevilla2019}%
  \BibitemOpen
  \bibfield  {author} {\bibinfo {author} {\bibfnamefont {F.~J.}\ \bibnamefont
  {Sevilla}}, \bibinfo {author} {\bibfnamefont {A.~V.}\ \bibnamefont {Arzola}},
  \ and\ \bibinfo {author} {\bibfnamefont {E.~P.}\ \bibnamefont {Cital}},\
  }\href {\doibase 10.1103/PhysRevE.99.012145} {\bibfield  {journal} {\bibinfo
  {journal} {Phys. Rev. E}\ }\textbf {\bibinfo {volume} {99}},\ \bibinfo
  {pages} {012145} (\bibinfo {year} {2019})}\BibitemShut {NoStop}%
\bibitem [{\citenamefont {Paksa}\ \emph {et~al.}(2016)\citenamefont {Paksa},
  \citenamefont {Bandemer}, \citenamefont {Hoeckendorf}, \citenamefont {Razin},
  \citenamefont {Tarbashevich}, \citenamefont {Minina}, \citenamefont {Meyen},
  \citenamefont {Biundo}, \citenamefont {Leidel}, \citenamefont {Peyrieras},
  \citenamefont {Gov}, \citenamefont {Keller},\ and\ \citenamefont
  {Raz}}]{Paksa2016}%
  \BibitemOpen
  \bibfield  {author} {\bibinfo {author} {\bibfnamefont {A.}~\bibnamefont
  {Paksa}}, \bibinfo {author} {\bibfnamefont {J.}~\bibnamefont {Bandemer}},
  \bibinfo {author} {\bibfnamefont {B.}~\bibnamefont {Hoeckendorf}}, \bibinfo
  {author} {\bibfnamefont {N.}~\bibnamefont {Razin}}, \bibinfo {author}
  {\bibfnamefont {K.}~\bibnamefont {Tarbashevich}}, \bibinfo {author}
  {\bibfnamefont {S.}~\bibnamefont {Minina}}, \bibinfo {author} {\bibfnamefont
  {D.}~\bibnamefont {Meyen}}, \bibinfo {author} {\bibfnamefont
  {A.}~\bibnamefont {Biundo}}, \bibinfo {author} {\bibfnamefont {S.~A.}\
  \bibnamefont {Leidel}}, \bibinfo {author} {\bibfnamefont {N.}~\bibnamefont
  {Peyrieras}}, \bibinfo {author} {\bibfnamefont {N.~S.}\ \bibnamefont {Gov}},
  \bibinfo {author} {\bibfnamefont {P.~J.}\ \bibnamefont {Keller}}, \ and\
  \bibinfo {author} {\bibfnamefont {E.}~\bibnamefont {Raz}},\ }\href {\doibase
  10.1038/ncomms11288} {\bibfield  {journal} {\bibinfo  {journal} {Nature
  Communications}\ }\textbf {\bibinfo {volume} {7}},\ \bibinfo {pages} {11288}
  (\bibinfo {year} {2016})}\BibitemShut {NoStop}%
\bibitem [{\citenamefont {Razin}\ \emph
  {et~al.}(2017{\natexlab{a}})\citenamefont {Razin}, \citenamefont {Voituriez},
  \citenamefont {Elgeti},\ and\ \citenamefont {Gov}}]{Razin2017}%
  \BibitemOpen
  \bibfield  {author} {\bibinfo {author} {\bibfnamefont {N.}~\bibnamefont
  {Razin}}, \bibinfo {author} {\bibfnamefont {R.}~\bibnamefont {Voituriez}},
  \bibinfo {author} {\bibfnamefont {J.}~\bibnamefont {Elgeti}}, \ and\ \bibinfo
  {author} {\bibfnamefont {N.~S.}\ \bibnamefont {Gov}},\ }\href {\doibase
  10.1103/PhysRevE.96.032606} {\bibfield  {journal} {\bibinfo  {journal} {Phys.
  Rev. E}\ }\textbf {\bibinfo {volume} {96}},\ \bibinfo {pages} {032606}
  (\bibinfo {year} {2017}{\natexlab{a}})}\BibitemShut {NoStop}%
\bibitem [{\citenamefont {Schnakenberg}(1976)}]{Schnakenberg1976}%
  \BibitemOpen
  \bibfield  {author} {\bibinfo {author} {\bibfnamefont {J.}~\bibnamefont
  {Schnakenberg}},\ }\href {\doibase 10.1103/RevModPhys.48.571} {\bibfield
  {journal} {\bibinfo  {journal} {Rev. Mod. Phys.}\ }\textbf {\bibinfo {volume}
  {48}},\ \bibinfo {pages} {571} (\bibinfo {year} {1976})}\BibitemShut
  {NoStop}%
\bibitem [{\citenamefont {Gaspard}(2004{\natexlab{a}})}]{Gaspard2004JCP}%
  \BibitemOpen
  \bibfield  {author} {\bibinfo {author} {\bibfnamefont {P.}~\bibnamefont
  {Gaspard}},\ }\href {\doibase 10.1063/1.1688758} {\bibfield  {journal}
  {\bibinfo  {journal} {The Journal of Chemical Physics}\ }\textbf {\bibinfo
  {volume} {120}},\ \bibinfo {pages} {8898} (\bibinfo {year}
  {2004}{\natexlab{a}})},\ \Eprint
  {http://arxiv.org/abs/https://doi.org/10.1063/1.1688758}
  {https://doi.org/10.1063/1.1688758} \BibitemShut {NoStop}%
\bibitem [{\citenamefont {Gaspard}(2004{\natexlab{b}})}]{Gaspard2004JSP}%
  \BibitemOpen
  \bibfield  {author} {\bibinfo {author} {\bibfnamefont {P.}~\bibnamefont
  {Gaspard}},\ }\href {\doibase 10.1007/s10955-004-3455-1} {\bibfield
  {journal} {\bibinfo  {journal} {Journal of Statistical Physics}\ }\textbf
  {\bibinfo {volume} {117}},\ \bibinfo {pages} {599} (\bibinfo {year}
  {2004}{\natexlab{b}})}\BibitemShut {NoStop}%
\bibitem [{\citenamefont {Lan}\ \emph {et~al.}(2012)\citenamefont {Lan},
  \citenamefont {Sartori}, \citenamefont {Neumann}, \citenamefont {Sourjik},\
  and\ \citenamefont {Tu}}]{Lan2012}%
  \BibitemOpen
  \bibfield  {author} {\bibinfo {author} {\bibfnamefont {G.}~\bibnamefont
  {Lan}}, \bibinfo {author} {\bibfnamefont {P.}~\bibnamefont {Sartori}},
  \bibinfo {author} {\bibfnamefont {S.}~\bibnamefont {Neumann}}, \bibinfo
  {author} {\bibfnamefont {V.}~\bibnamefont {Sourjik}}, \ and\ \bibinfo
  {author} {\bibfnamefont {Y.}~\bibnamefont {Tu}},\ }\href {\doibase
  10.1038/nphys2276} {\bibfield  {journal} {\bibinfo  {journal} {Nature
  Physics}\ }\textbf {\bibinfo {volume} {8}},\ \bibinfo {pages} {422} (\bibinfo
  {year} {2012})}\BibitemShut {NoStop}%
\bibitem [{\citenamefont {Andrae}\ \emph {et~al.}(2010)\citenamefont {Andrae},
  \citenamefont {Cremer}, \citenamefont {Reichenbach},\ and\ \citenamefont
  {Frey}}]{Benjamin2010}%
  \BibitemOpen
  \bibfield  {author} {\bibinfo {author} {\bibfnamefont {B.}~\bibnamefont
  {Andrae}}, \bibinfo {author} {\bibfnamefont {J.}~\bibnamefont {Cremer}},
  \bibinfo {author} {\bibfnamefont {T.}~\bibnamefont {Reichenbach}}, \ and\
  \bibinfo {author} {\bibfnamefont {E.}~\bibnamefont {Frey}},\ }\href {\doibase
  10.1103/PhysRevLett.104.218102} {\bibfield  {journal} {\bibinfo  {journal}
  {Phys. Rev. Lett.}\ }\textbf {\bibinfo {volume} {104}},\ \bibinfo {pages}
  {218102} (\bibinfo {year} {2010})}\BibitemShut {NoStop}%
\bibitem [{\citenamefont {Li}\ \emph {et~al.}(2019)\citenamefont {Li},
  \citenamefont {Horowitz}, \citenamefont {Gingrich},\ and\ \citenamefont
  {Fakhri}}]{Li2019}%
  \BibitemOpen
  \bibfield  {author} {\bibinfo {author} {\bibfnamefont {J.}~\bibnamefont
  {Li}}, \bibinfo {author} {\bibfnamefont {J.~M.}\ \bibnamefont {Horowitz}},
  \bibinfo {author} {\bibfnamefont {T.~R.}\ \bibnamefont {Gingrich}}, \ and\
  \bibinfo {author} {\bibfnamefont {N.}~\bibnamefont {Fakhri}},\ }\href
  {\doibase 10.1038/s41467-019-09631-x} {\bibfield  {journal} {\bibinfo
  {journal} {Nature Communications}\ }\textbf {\bibinfo {volume} {10}},\
  \bibinfo {pages} {1666} (\bibinfo {year} {2019})}\BibitemShut {NoStop}%
\bibitem [{\citenamefont {Busiello}\ and\ \citenamefont
  {Maritan}(2019)}]{Busiello2019}%
  \BibitemOpen
  \bibfield  {author} {\bibinfo {author} {\bibfnamefont {D.~M.}\ \bibnamefont
  {Busiello}}\ and\ \bibinfo {author} {\bibfnamefont {A.}~\bibnamefont
  {Maritan}},\ }\href {\doibase 10.1088/1742-5468/ab363e} {\bibfield  {journal}
  {\bibinfo  {journal} {Journal of Statistical Mechanics: Theory and
  Experiment}\ }\textbf {\bibinfo {volume} {2019}},\ \bibinfo {pages} {104013}
  (\bibinfo {year} {2019})}\BibitemShut {NoStop}%
\bibitem [{\citenamefont {Ganguly}\ and\ \citenamefont
  {Chaudhuri}(2013)}]{Ganguly2013}%
  \BibitemOpen
  \bibfield  {author} {\bibinfo {author} {\bibfnamefont {C.}~\bibnamefont
  {Ganguly}}\ and\ \bibinfo {author} {\bibfnamefont {D.}~\bibnamefont
  {Chaudhuri}},\ }\href {\doibase 10.1103/PhysRevE.88.032102} {\bibfield
  {journal} {\bibinfo  {journal} {Phys. Rev. E}\ }\textbf {\bibinfo {volume}
  {88}},\ \bibinfo {pages} {032102} (\bibinfo {year} {2013})}\BibitemShut
  {NoStop}%
\bibitem [{\citenamefont {Chaudhuri}(2014)}]{Chaudhuri2014}%
  \BibitemOpen
  \bibfield  {author} {\bibinfo {author} {\bibfnamefont {D.}~\bibnamefont
  {Chaudhuri}},\ }\href {\doibase 10.1103/PhysRevE.90.022131} {\bibfield
  {journal} {\bibinfo  {journal} {Phys. Rev. E}\ }\textbf {\bibinfo {volume}
  {90}},\ \bibinfo {pages} {022131} (\bibinfo {year} {2014})}\BibitemShut
  {NoStop}%
\bibitem [{\citenamefont {Fodor}\ \emph {et~al.}(2016)\citenamefont {Fodor},
  \citenamefont {Nardini}, \citenamefont {Cates}, \citenamefont {Tailleur},
  \citenamefont {Visco},\ and\ \citenamefont {van Wijland}}]{Fodor2016}%
  \BibitemOpen
  \bibfield  {author} {\bibinfo {author} {\bibfnamefont {E.}~\bibnamefont
  {Fodor}}, \bibinfo {author} {\bibfnamefont {C.}~\bibnamefont {Nardini}},
  \bibinfo {author} {\bibfnamefont {M.~E.}\ \bibnamefont {Cates}}, \bibinfo
  {author} {\bibfnamefont {J.}~\bibnamefont {Tailleur}}, \bibinfo {author}
  {\bibfnamefont {P.}~\bibnamefont {Visco}}, \ and\ \bibinfo {author}
  {\bibfnamefont {F.}~\bibnamefont {van Wijland}},\ }\href {\doibase
  10.1103/PhysRevLett.117.038103} {\bibfield  {journal} {\bibinfo  {journal}
  {Phys. Rev. Lett.}\ }\textbf {\bibinfo {volume} {117}},\ \bibinfo {pages}
  {038103} (\bibinfo {year} {2016})}\BibitemShut {NoStop}%
\bibitem [{\citenamefont {Falasco}\ \emph {et~al.}(2016)\citenamefont
  {Falasco}, \citenamefont {Pfaller}, \citenamefont {Bregulla}, \citenamefont
  {Cichos},\ and\ \citenamefont {Kroy}}]{Falasco2016}%
  \BibitemOpen
  \bibfield  {author} {\bibinfo {author} {\bibfnamefont {G.}~\bibnamefont
  {Falasco}}, \bibinfo {author} {\bibfnamefont {R.}~\bibnamefont {Pfaller}},
  \bibinfo {author} {\bibfnamefont {A.~P.}\ \bibnamefont {Bregulla}}, \bibinfo
  {author} {\bibfnamefont {F.}~\bibnamefont {Cichos}}, \ and\ \bibinfo {author}
  {\bibfnamefont {K.}~\bibnamefont {Kroy}},\ }\href {\doibase
  10.1103/PhysRevE.94.030602} {\bibfield  {journal} {\bibinfo  {journal} {Phys.
  Rev. E}\ }\textbf {\bibinfo {volume} {94}},\ \bibinfo {pages} {030602}
  (\bibinfo {year} {2016})}\BibitemShut {NoStop}%
\bibitem [{\citenamefont {Nardini}\ \emph {et~al.}(2017)\citenamefont
  {Nardini}, \citenamefont {Fodor}, \citenamefont {Tjhung}, \citenamefont {van
  Wijland}, \citenamefont {Tailleur},\ and\ \citenamefont
  {Cates}}]{Nardini2017}%
  \BibitemOpen
  \bibfield  {author} {\bibinfo {author} {\bibfnamefont {C.}~\bibnamefont
  {Nardini}}, \bibinfo {author} {\bibfnamefont {E.}~\bibnamefont {Fodor}},
  \bibinfo {author} {\bibfnamefont {E.}~\bibnamefont {Tjhung}}, \bibinfo
  {author} {\bibfnamefont {F.}~\bibnamefont {van Wijland}}, \bibinfo {author}
  {\bibfnamefont {J.}~\bibnamefont {Tailleur}}, \ and\ \bibinfo {author}
  {\bibfnamefont {M.~E.}\ \bibnamefont {Cates}},\ }\href {\doibase
  10.1103/PhysRevX.7.021007} {\bibfield  {journal} {\bibinfo  {journal} {Phys.
  Rev. X}\ }\textbf {\bibinfo {volume} {7}},\ \bibinfo {pages} {021007}
  (\bibinfo {year} {2017})}\BibitemShut {NoStop}%
\bibitem [{\citenamefont {Mandal}\ \emph {et~al.}(2017)\citenamefont {Mandal},
  \citenamefont {Klymko},\ and\ \citenamefont {DeWeese}}]{Mandal2017}%
  \BibitemOpen
  \bibfield  {author} {\bibinfo {author} {\bibfnamefont {D.}~\bibnamefont
  {Mandal}}, \bibinfo {author} {\bibfnamefont {K.}~\bibnamefont {Klymko}}, \
  and\ \bibinfo {author} {\bibfnamefont {M.~R.}\ \bibnamefont {DeWeese}},\
  }\href {\doibase 10.1103/PhysRevLett.119.258001} {\bibfield  {journal}
  {\bibinfo  {journal} {Phys. Rev. Lett.}\ }\textbf {\bibinfo {volume} {119}},\
  \bibinfo {pages} {258001} (\bibinfo {year} {2017})}\BibitemShut {NoStop}%
\bibitem [{\citenamefont {Marconi}\ \emph {et~al.}(2017)\citenamefont
  {Marconi}, \citenamefont {Puglisi},\ and\ \citenamefont
  {Maggi}}]{Marconi2017}%
  \BibitemOpen
  \bibfield  {author} {\bibinfo {author} {\bibfnamefont {U.~M.~B.}\
  \bibnamefont {Marconi}}, \bibinfo {author} {\bibfnamefont {A.}~\bibnamefont
  {Puglisi}}, \ and\ \bibinfo {author} {\bibfnamefont {C.}~\bibnamefont
  {Maggi}},\ }\href {\doibase 10.1038/srep46496} {\bibfield  {journal}
  {\bibinfo  {journal} {Scientific Reports}\ }\textbf {\bibinfo {volume} {7}},\
  \bibinfo {pages} {46496} (\bibinfo {year} {2017})}\BibitemShut {NoStop}%
\bibitem [{\citenamefont {Pietzonka}\ and\ \citenamefont
  {Seifert}(2017)}]{Pietzonka2017}%
  \BibitemOpen
  \bibfield  {author} {\bibinfo {author} {\bibfnamefont {P.}~\bibnamefont
  {Pietzonka}}\ and\ \bibinfo {author} {\bibfnamefont {U.}~\bibnamefont
  {Seifert}},\ }\href {\doibase 10.1088/1751-8121/aa91b9} {\bibfield  {journal}
  {\bibinfo  {journal} {Journal of Physics A: Mathematical and Theoretical}\
  }\textbf {\bibinfo {volume} {51}},\ \bibinfo {pages} {01LT01} (\bibinfo
  {year} {2017})}\BibitemShut {NoStop}%
\bibitem [{\citenamefont {Shankar}\ and\ \citenamefont
  {Marchetti}(2018)}]{Shankar2018}%
  \BibitemOpen
  \bibfield  {author} {\bibinfo {author} {\bibfnamefont {S.}~\bibnamefont
  {Shankar}}\ and\ \bibinfo {author} {\bibfnamefont {M.~C.}\ \bibnamefont
  {Marchetti}},\ }\href {\doibase 10.1103/PhysRevE.98.020604} {\bibfield
  {journal} {\bibinfo  {journal} {Phys. Rev. E}\ }\textbf {\bibinfo {volume}
  {98}},\ \bibinfo {pages} {020604} (\bibinfo {year} {2018})}\BibitemShut
  {NoStop}%
\bibitem [{\citenamefont {Speck}(2018)}]{Speck2018}%
  \BibitemOpen
  \bibfield  {author} {\bibinfo {author} {\bibfnamefont {T.}~\bibnamefont
  {Speck}},\ }\href {\doibase 10.1209/0295-5075/123/20007} {\bibfield
  {journal} {\bibinfo  {journal} {{EPL} (Europhysics Letters)}\ }\textbf
  {\bibinfo {volume} {123}},\ \bibinfo {pages} {20007} (\bibinfo {year}
  {2018})}\BibitemShut {NoStop}%
\bibitem [{\citenamefont {Caprini}\ \emph {et~al.}(2019)\citenamefont
  {Caprini}, \citenamefont {Marconi}, \citenamefont {Puglisi},\ and\
  \citenamefont {Vulpiani}}]{Caprini2019}%
  \BibitemOpen
  \bibfield  {author} {\bibinfo {author} {\bibfnamefont {L.}~\bibnamefont
  {Caprini}}, \bibinfo {author} {\bibfnamefont {U.~M.~B.}\ \bibnamefont
  {Marconi}}, \bibinfo {author} {\bibfnamefont {A.}~\bibnamefont {Puglisi}}, \
  and\ \bibinfo {author} {\bibfnamefont {A.}~\bibnamefont {Vulpiani}},\ }\href
  {\doibase 10.1088/1742-5468/ab14dd} {\bibfield  {journal} {\bibinfo
  {journal} {Journal of Statistical Mechanics: Theory and Experiment}\ }\textbf
  {\bibinfo {volume} {2019}},\ \bibinfo {pages} {053203} (\bibinfo {year}
  {2019})}\BibitemShut {NoStop}%
\bibitem [{\citenamefont {Dabelow}\ \emph {et~al.}(2019)\citenamefont
  {Dabelow}, \citenamefont {Bo},\ and\ \citenamefont {Eichhorn}}]{Dabelow2019}%
  \BibitemOpen
  \bibfield  {author} {\bibinfo {author} {\bibfnamefont {L.}~\bibnamefont
  {Dabelow}}, \bibinfo {author} {\bibfnamefont {S.}~\bibnamefont {Bo}}, \ and\
  \bibinfo {author} {\bibfnamefont {R.}~\bibnamefont {Eichhorn}},\ }\href
  {\doibase 10.1103/PhysRevX.9.021009} {\bibfield  {journal} {\bibinfo
  {journal} {Phys. Rev. X}\ }\textbf {\bibinfo {volume} {9}},\ \bibinfo {pages}
  {021009} (\bibinfo {year} {2019})}\BibitemShut {NoStop}%
\bibitem [{\citenamefont {Szamel}(2019)}]{Szamel2019}%
  \BibitemOpen
  \bibfield  {author} {\bibinfo {author} {\bibfnamefont {G.}~\bibnamefont
  {Szamel}},\ }\href {\doibase 10.1103/PhysRevE.100.050603} {\bibfield
  {journal} {\bibinfo  {journal} {Phys. Rev. E}\ }\textbf {\bibinfo {volume}
  {100}},\ \bibinfo {pages} {050603} (\bibinfo {year} {2019})}\BibitemShut
  {NoStop}%
\bibitem [{\citenamefont {Flenner}\ and\ \citenamefont
  {Szamel}(2020)}]{Flenner2020}%
  \BibitemOpen
  \bibfield  {author} {\bibinfo {author} {\bibfnamefont {E.}~\bibnamefont
  {Flenner}}\ and\ \bibinfo {author} {\bibfnamefont {G.}~\bibnamefont
  {Szamel}},\ }\href {\doibase 10.1103/PhysRevE.102.022607} {\bibfield
  {journal} {\bibinfo  {journal} {Phys. Rev. E}\ }\textbf {\bibinfo {volume}
  {102}},\ \bibinfo {pages} {022607} (\bibinfo {year} {2020})}\BibitemShut
  {NoStop}%
\bibitem [{\citenamefont {Chaki}\ and\ \citenamefont
  {Chakrabarti}(2018)}]{Chaki2018}%
  \BibitemOpen
  \bibfield  {author} {\bibinfo {author} {\bibfnamefont {S.}~\bibnamefont
  {Chaki}}\ and\ \bibinfo {author} {\bibfnamefont {R.}~\bibnamefont
  {Chakrabarti}},\ }\href {\doibase
  https://doi.org/10.1016/j.physa.2018.07.055} {\bibfield  {journal} {\bibinfo
  {journal} {Physica A: Statistical Mechanics and its Applications}\ }\textbf
  {\bibinfo {volume} {511}},\ \bibinfo {pages} {302 } (\bibinfo {year}
  {2018})}\BibitemShut {NoStop}%
\bibitem [{\citenamefont {Chaki}\ and\ \citenamefont
  {Chakrabarti}(2019)}]{Chaki2019}%
  \BibitemOpen
  \bibfield  {author} {\bibinfo {author} {\bibfnamefont {S.}~\bibnamefont
  {Chaki}}\ and\ \bibinfo {author} {\bibfnamefont {R.}~\bibnamefont
  {Chakrabarti}},\ }\href {\doibase
  https://doi.org/10.1016/j.physa.2019.121574} {\bibfield  {journal} {\bibinfo
  {journal} {Physica A: Statistical Mechanics and its Applications}\ }\textbf
  {\bibinfo {volume} {530}},\ \bibinfo {pages} {121574} (\bibinfo {year}
  {2019})}\BibitemShut {NoStop}%
\bibitem [{\citenamefont {GrandPre}\ \emph {et~al.}(2020)\citenamefont
  {GrandPre}, \citenamefont {Klymko}, \citenamefont {Mandadapu},\ and\
  \citenamefont {Limmer}}]{gr2020entropy}%
  \BibitemOpen
  \bibfield  {author} {\bibinfo {author} {\bibfnamefont {T.}~\bibnamefont
  {GrandPre}}, \bibinfo {author} {\bibfnamefont {K.}~\bibnamefont {Klymko}},
  \bibinfo {author} {\bibfnamefont {K.~K.}\ \bibnamefont {Mandadapu}}, \ and\
  \bibinfo {author} {\bibfnamefont {D.~T.}\ \bibnamefont {Limmer}},\
  }\href@noop {} {\enquote {\bibinfo {title} {Entropy production fluctuations
  encode collective behavior in active matter},}\ } (\bibinfo {year} {2020}),\
  \Eprint {http://arxiv.org/abs/2007.12149} {arXiv:2007.12149
  [cond-mat.stat-mech]} \BibitemShut {NoStop}%
\bibitem [{\citenamefont {Caballero}\ and\ \citenamefont
  {Cates}(2020)}]{Caballero2020}%
  \BibitemOpen
  \bibfield  {author} {\bibinfo {author} {\bibfnamefont {F.}~\bibnamefont
  {Caballero}}\ and\ \bibinfo {author} {\bibfnamefont {M.~E.}\ \bibnamefont
  {Cates}},\ }\href {\doibase 10.1103/PhysRevLett.124.240604} {\bibfield
  {journal} {\bibinfo  {journal} {Phys. Rev. Lett.}\ }\textbf {\bibinfo
  {volume} {124}},\ \bibinfo {pages} {240604} (\bibinfo {year}
  {2020})}\BibitemShut {NoStop}%
\bibitem [{\citenamefont {Razin}\ \emph
  {et~al.}(2017{\natexlab{b}})\citenamefont {Razin}, \citenamefont {Voituriez},
  \citenamefont {Elgeti},\ and\ \citenamefont {Gov}}]{Razin2017b}%
  \BibitemOpen
  \bibfield  {author} {\bibinfo {author} {\bibfnamefont {N.}~\bibnamefont
  {Razin}}, \bibinfo {author} {\bibfnamefont {R.}~\bibnamefont {Voituriez}},
  \bibinfo {author} {\bibfnamefont {J.}~\bibnamefont {Elgeti}}, \ and\ \bibinfo
  {author} {\bibfnamefont {N.~S.}\ \bibnamefont {Gov}},\ }\href {\doibase
  10.1103/PhysRevE.96.052409} {\bibfield  {journal} {\bibinfo  {journal} {Phys.
  Rev. E}\ }\textbf {\bibinfo {volume} {96}},\ \bibinfo {pages} {052409}
  (\bibinfo {year} {2017}{\natexlab{b}})}\BibitemShut {NoStop}%
\bibitem [{\citenamefont {Marchetti}\ \emph {et~al.}(2016)\citenamefont
  {Marchetti}, \citenamefont {Fily}, \citenamefont {Henkes}, \citenamefont
  {Patch},\ and\ \citenamefont {Yllanes}}]{Marchetti2016}%
  \BibitemOpen
  \bibfield  {author} {\bibinfo {author} {\bibfnamefont {M.~C.}\ \bibnamefont
  {Marchetti}}, \bibinfo {author} {\bibfnamefont {Y.}~\bibnamefont {Fily}},
  \bibinfo {author} {\bibfnamefont {S.}~\bibnamefont {Henkes}}, \bibinfo
  {author} {\bibfnamefont {A.}~\bibnamefont {Patch}}, \ and\ \bibinfo {author}
  {\bibfnamefont {D.}~\bibnamefont {Yllanes}},\ }\href {\doibase
  http://dx.doi.org/10.1016/j.cocis.2016.01.003} {\bibfield  {journal}
  {\bibinfo  {journal} {Current Opinion in Colloid \& Interface Science}\
  }\textbf {\bibinfo {volume} {21}},\ \bibinfo {pages} {34} (\bibinfo {year}
  {2016})}\BibitemShut {NoStop}%
\bibitem [{\citenamefont {Cates}\ and\ \citenamefont
  {Tailleur}(2015)}]{Cates2015}%
  \BibitemOpen
  \bibfield  {author} {\bibinfo {author} {\bibfnamefont {M.~E.}\ \bibnamefont
  {Cates}}\ and\ \bibinfo {author} {\bibfnamefont {J.}~\bibnamefont
  {Tailleur}},\ }\href {\doibase 10.1146/annurev-conmatphys-031214-014710}
  {\bibfield  {journal} {\bibinfo  {journal} {Annual Review of Condensed Matter
  Physics}\ }\textbf {\bibinfo {volume} {6}},\ \bibinfo {pages} {219} (\bibinfo
  {year} {2015})}\BibitemShut {NoStop}%
\bibitem [{\citenamefont {Marini Bettolo~Marconi}\ and\ \citenamefont
  {Maggi}(2015)}]{Marconi2015}%
  \BibitemOpen
  \bibfield  {author} {\bibinfo {author} {\bibfnamefont {U.}~\bibnamefont
  {Marini Bettolo~Marconi}}\ and\ \bibinfo {author} {\bibfnamefont
  {C.}~\bibnamefont {Maggi}},\ }\href {\doibase 10.1039/C5SM01718A} {\bibfield
  {journal} {\bibinfo  {journal} {Soft Matter}\ }\textbf {\bibinfo {volume}
  {11}},\ \bibinfo {pages} {8768} (\bibinfo {year} {2015})}\BibitemShut
  {NoStop}%
\bibitem [{\citenamefont {Tom{\'e}}(2006)}]{Tome2006}%
  \BibitemOpen
  \bibfield  {author} {\bibinfo {author} {\bibfnamefont {T.}~\bibnamefont
  {Tom{\'e}}},\ }\href
  {http://www.scielo.br/scielo.php?script=sci_arttext&pid=S0103-97332006000700029&nrm=iso}
  {\bibfield  {journal} {\bibinfo  {journal} {{Brazilian Journal of Physics}}\
  }\textbf {\bibinfo {volume} {36}},\ \bibinfo {pages} {1285 } (\bibinfo {year}
  {2006})}\BibitemShut {NoStop}%
\bibitem [{\citenamefont {Seifert}(2005)}]{Seifert2005}%
  \BibitemOpen
  \bibfield  {author} {\bibinfo {author} {\bibfnamefont {U.}~\bibnamefont
  {Seifert}},\ }\href {\doibase 10.1103/PhysRevLett.95.040602} {\bibfield
  {journal} {\bibinfo  {journal} {Phys. Rev. Lett.}\ }\textbf {\bibinfo
  {volume} {95}},\ \bibinfo {pages} {040602} (\bibinfo {year}
  {2005})}\BibitemShut {NoStop}%
\bibitem [{\citenamefont {Tom\'e}\ and\ \citenamefont
  {de~Oliveira}(2012)}]{Tome2012}%
  \BibitemOpen
  \bibfield  {author} {\bibinfo {author} {\bibfnamefont {T.}~\bibnamefont
  {Tom\'e}}\ and\ \bibinfo {author} {\bibfnamefont {M.~J.}\ \bibnamefont
  {de~Oliveira}},\ }\href {\doibase 10.1103/PhysRevLett.108.020601} {\bibfield
  {journal} {\bibinfo  {journal} {Phys. Rev. Lett.}\ }\textbf {\bibinfo
  {volume} {108}},\ \bibinfo {pages} {020601} (\bibinfo {year}
  {2012})}\BibitemShut {NoStop}%
\end{thebibliography}
%

\end{document}